%% file: main.tex
\documentclass[a4paper,journal,twoside]{IEEEtran}
\IEEEoverridecommandlockouts

\newcommand\hmmax{0}
\newcommand\bmmax{0}

\usepackage{graphicx}
\usepackage{tikz}
\usepackage{pgfplots}
\usepackage{xcolor}
\usepackage{color, colortbl}
\usepackage{amsmath}
\usepackage{bbm}
\usepackage{amsfonts, amssymb}
\usepackage{mathtools}
\usepackage{enumerate}
\usepackage{hyperref} 		
\usepackage{cite}
\usepackage[nolist]{acronym}
\usepackage{amsmath}

\usepackage{booktabs} 		
\usepackage{diagbox}		
\usepackage{upgreek}
\usepackage{bbm}
\usepackage{Files/bm}
\usepackage{Files/winsnotation}
\usepackage{Files/Symbols_OPL}
\usepackage{Files/LVcolours} 

\usepackage[ruled,vlined,linesnumbered]{algorithm2e} 
\usepackage{setspace}	 	

\newcommand{\InlineIfThen}[2]{\textbf{if}~#1~\textbf{then}~#2}

\usepackage{comment}
\usepackage{physics}
\usepackage{enumitem}

\newlist{myitemize}{itemize}{3}
\setlist[myitemize,1]{label=1.,leftmargin=1em}
\setlist[myitemize,2]{label=$\rightarrow$,leftmargin=0.75em}
\setlist[myitemize,3]{label=$\diamond$}

\usepackage{array}
\newcolumntype{C}[1]{>{\centering\arraybackslash}p{#1}}

\makeatletter
\def\endthebibliography{%
  \def\@noitemerr{\@latex@warning{Empty `thebibliography' environment}}%
  \endlist
}
\makeatother
\usepackage{amsthm}
\theoremstyle{definition}

\usepackage{algpseudocode,amsmath}

\usepackage{multirow}

\pgfplotsset{compat=1.17}

\begin{document}

\title{Restart Belief: A General Quantum LDPC Decoder}

\author{Lorenzo~Valentini,~\IEEEmembership{Member,~IEEE,} 
Diego Forlivesi,~\IEEEmembership{Graduate~Student~Member,~IEEE,}
Andrea Talarico, 
        and~Marco~Chiani,~\IEEEmembership{Fellow,~IEEE}
\thanks{The authors are with the Department of Electrical, Electronic, and Information Engineering ``Guglielmo Marconi'' and CNIT/WiLab, University of Bologna, V.le Risorgimento 2, 40136 Bologna, Italy. E-mail: \{lorenzo.valentini13, diego.forlivesi2, andrea.talarico3, marco.chiani\}@unibo.it. 
}
}

\maketitle 

\input{Files/Acronimi_SICMMA.tex}
\setcounter{page}{1}

\begin{abstract}
Hardware-friendly quantum low-density parity-check (QLDPC) decoders are commonly built upon belief propagation (BP) processing.
Yet, quantum degeneracy often prevents BP from achieving reliable convergence. 
To overcome this fundamental limitation, we propose the restart belief (RB) decoder, an iterative BP-based algorithm inspired by branch-and-bound optimization principles.
From our analysis we find that the RB decoder represents both the fastest and most accurate decoding algorithm applicable to QLDPC codes to date, conceived with the explicit goal of approaching error correction up to the code distance.
\end{abstract}


\section{Introduction}

Preserving quantum information through \ac{QEC} is essential for scalable quantum computing and reliable quantum memories \cite{Sho:95}.
Because quantum states are extremely fragile, \ac{QEC} encodes logical qubits redundantly into many physical qubits and continuously detects and corrects errors without collapsing the state.
In a typical \ac{QEC} cycle, the system periodically measures stabilizer generators, identifies potential errors through syndrome extraction, and applies corrective operations to restore the quantum information~\cite{Got:96}. 
Therefore, achieving practical fault-tolerant operation requires decoders that are not only accurate, but also extremely fast and hardware-friendly, capable of running efficiently on specialized platforms such as \acp{FPGA} or \acp{ASIC}~\cite{Mul25:relay}.

Among various \ac{QEC} schemes, \ac{QLDPC} codes are particularly promising due to their low-weight stabilizer generators, which reduce circuit complexity and error propagation while maintaining good distance properties.
In particular, the \ac{CSS} version of these codes is advantageous, as the separation between $\M{Z}$-type and $\M{X}$-type generators simplifies both the circuit design and the decoding process, while enabling many transversal encoded operations that are highly desirable for fault-tolerant quantum computation~\cite{CalSho:96, Ste:96}.
These features make them attractive for large-scale fault-tolerant architectures where both qubit efficiency and reliable error suppression are critical.
In this context, surface codes \cite{BraKit:98}, structured instances of \ac{QLDPC} codes, have received significant attention, with several specialized decoders such as \ac{MWPM}, \ac{UF}, and \ac{BC} \cite{HigGid:23, Del:21, ForValChi:24STM, ForValChi25:Bub}.
However, these decoders are highly specialized and tailored to the specific structure of surface codes.
For this reason, researchers have actively explored \ac{BP}-based approaches for decoding general \ac{QLDPC} codes.

In this direction, \cite{PanKalOSD:21} proposed enhancing the \ac{BP} algorithm by incorporating a post-processing stage based on \ac{OSD}, resulting in the combined decoding strategy known as \ac{BP}+\ac{OSD}.
However, this approach has certain drawbacks: the \ac{OSD} step requires matrix inversion, which can be computationally expensive, and it does not always preserve the code distance.
Subsequently, \cite{Yao24:GD} introduced the decoder \ac{BPGD}, in which multiple instances of \ac{BP} are executed iteratively until convergence is achieved.
In each iteration where the decoder becomes trapped, the most reliable \ac{LLR} is fixed to guide the decoding process toward convergence.
This algorithm is computationally more efficient, however, it provides a limited performance advantage compared to conventional \ac{BP}.
Recently, \cite{Mul25:relay} introduced a novel decoding scheme, termed \ac{RelayBP}, which aims to simultaneously achieve high decoding performance and low latency. 
The core concept involves running the \ac{BP} algorithm and, in cases where convergence is not reached, restarting the process while updating the input \acp{LLR} using a damped version of the output \acp{LLR} from the previous iteration. 
This decoder has been shown to obtain very good performance and decoding complexity for certain \ac{QLDPC} codes such as the bivariate bicycle and the rotated surface ones.

In this paper, we continue the search for a general \ac{QLDPC} decoder that balances performance and complexity.
Our proposed decoder, hereafter named \ac{RB}, executes a first \ac{BP} and its output \acp{LLR} are sorted. 
The minimum \ac{LLR} is used to guide the decoder by applying an error to the corresponding qubit and updating the syndrome and the qubit \ac{LLR} before the next \ac{BP}.
If the decoder still fails to converge, this process is repeated.
Once convergence is achieved and a valid solution is obtained, the process is restarted using the second smallest \ac{LLR} from the root \ac{BP}, then the third, and so forth until a maximum is reached.
In this way, the minimum \acp{LLR} of the root generate each a solution exploring a branch of the decision tree.
To enhance the speed of the decoder we also include code distance-specific and defect-specific early termination strategies.
In addition, it is worth noting that all the branches \ac{BP}, generating a solution, can be efficiently executed in parallel.
To benchmark and validate our \ac{RB} decoder we selected various \ac{QLDPC} codes and compare the results with all the above mentioned \ac{QLDPC} decoders based on \ac{BP}.

\section{Efficient Decoders for QLDPC Codes}
\label{sec:decodersLiterature}

The Pauli operators are denoted by $\M{X}$, $\M{Y}$, and $\M{Z}$.
A quantum error correcting code encoding $k$ logical qubits $\ket{\varphi}$ into $n$ physical qubits $\ket{\psi}$ with minimum distance $d$ is represented as $[[n,k,d]]$, and can correct any error affecting up to $t = \lfloor (d-1)/2 \rfloor$ qubits.
In the stabilizer formalism, the code is defined by $n-k$ independent, commuting generators $\M{G}_i \in \mathcal{G}_n$, where $\mathcal{G}_n$ denotes the $n$-qubit Pauli group~\cite{Got:09}.
The subgroup generated by all $\M{G}_i$ forms the stabilizer $\mathcal{S}$, and the code space $\mathcal{C}$ consists of all states $\ket{\psi}$ satisfying $\M{S}\ket{\psi}=\ket{\psi}$ for every $\M{S} \in \mathcal{S}$.
Stabilizer generators correspond to measurements that leave the encoded state unchanged and are implemented using ancillary qubits.
When an error $\M{E} \in \mathcal{G}_n$ acts on a codeword, producing $\M{E}\ket{\psi}$, the resulting binary syndrome vector $\V{s}$ encodes commutation information: $s_i = 0$ if $\M{G}_i$ commutes with $\M{E}$ and $s_i = 1$ otherwise.

In this context, the structure of a code can be represented as a bipartite graph, where two types of nodes are distinguished: variable nodes and check nodes. 
Variable nodes correspond to the errors on data qubits, while check nodes represent the generator constraints. 
The goal of decoding is to determine the most likely configuration of errors on the variable nodes that is consistent with the observed syndrome.
In the following, we consider \ac{CSS} codes \cite{CalSho:96,Ste:96}, where the decoding of $\M{X}$ and $\M{Z}$ Pauli errors can be treated independently. 
Therefore, for the sake of presentation, we focus on the correction of $\M{Z}$ Pauli errors and consider as check nodes only the $\M{X}$ stabilizer generators. 
Accordingly, the parity check matrix $\M{H}_\mathrm{x}$ refers to the matrix constructed from these $\M{X}$ generators, with each row corresponding to one generator. To simplify notation, we denote the syndrome of the current decoder by $\V{s}$.
The decoding of 
$\M{X}$ errors can be performed in an analogous manner.

\subsection{Belief Propagation}

The \ac{BP} algorithm iteratively exchanges probability messages between check and variable nodes to infer the locations of the most likely errors which are consistent with the measured syndrome. 
We denote as $m_{u_i \rightarrow v_j}$ the message from check node $u_i$ to variable node $v_j$, and as $m_{v_j \rightarrow u_i}$ the message from a variable node $v_j$ to a check node $u_i$.
Also, we denote by $\mathcal{N}(u_i)$ the set of all neighboring variable nodes connected to the check node $u_i$. 
Analogously, $\mathcal{N}(v_j)$ is the set of check nodes neighboring the variable node $v_j$.

Given a check node $i$ and a variable node $j$, the check-to-variable messages according to the scaled min-sum approach are computed as~ \cite{CheFos02:alpha}
\begin{align}
\label{eq:BP_CtoV}
m_{u_i \rightarrow v_j} 
&= (-1)^{s_i}\alpha
\!\left[\prod_{v' \in \mathcal{N}(u_i)\setminus v_j}\!\! 
\operatorname{sign}\!\big(m_{v' \rightarrow u_i}\big)\right] \notag\\
&\quad \min_{v' \in \mathcal{N}(u_i)\setminus v_j}\!\! 
\big|m_{v' \rightarrow u_i}\big|,
\end{align}
where $\alpha$ is a scaling factor, and $s_i$ represents the $i$-th syndrome bit.
Similarly, the variable-to-check messages
are computed as~\cite{CheFos02:alpha}
\begin{align}
\label{eq:BP_VtoC}
m_{v_j \rightarrow u_i} 
&= p_j^{\text{in}} + \sum_{u' \in \mathcal{N}(v_j)\setminus u_i} m_{u' \rightarrow v_j}
\end{align}
where $p_j^{\text{in}}$ is the input \ac{LLR} computed with the channel prior information as $p_j^{\text{in}}=\log((1-p)/p)$, with $p$ the physical error rate of the channel.

Next, the output \ac{LLR} for each variable node is~\cite{CheFos02:alpha}
\begin{align}
\label{eq:finalLLR}
p_j^{\text{out}} &= p_j^{\text{in}} + \sum_{u' \in \mathcal{N}(v_j)} m_{u' \rightarrow v_j}.
\end{align}
Finally, a hard decision is applied to determine a likely error as $\hat{e}_j = \frac{1}{2} - \frac{1}{2}\operatorname{sign}(p_j^{\text{out}})$, which is verified to ensure that the proposed error configuration is consistent with the observed syndrome.
In the following we denote as $T$ the number of \ac{BP} iterations, i.e., the number of times \eqref{eq:BP_CtoV} and \eqref{eq:BP_VtoC} are iterated.

\subsection{Ordered Statistic}

The \ac{BP} algorithm alone can fail to converge for quantum codes when multiple degenerate error configurations are equally likely, causing the decoder to assign significant probability to more than one error pattern simultaneously \cite{PouYeo08:QLDPC_BP}.
To address this, \ac{OSD} is applied as a post-processing step, refining the BP estimate to produce a syndrome-compliant error configuration and ensuring reliable convergence~\cite{PanKalOSD:21}.
This algorithm leverages the \ac{BP} output \ac{LLR} to construct an invertible, full-rank matrix $\M{H}_s$ of size $r \times r$, which is then used to recover an estimated error vector $\hat{\mathbf{e}}_s \in \mathbb{F}_2^r$ from the syndrome $\mathbf{s}$ via $\hat{\mathbf{e}}_s = \M{H}_s^{-1} \mathbf{s}$.
The remaining qubit errors, associated to the vector $\hat{\mathbf{e}}_t \in \mathbb{F}_2^{\,n-r}$, are set to $\mathbf{0}$.
The matrix $\M{H}_s$ is constructed by first reordering the columns of the parity-check matrix $\M{H}_\mathrm{x}$ according to the estimated probability of error on each qubit, starting with the most likely erroneous qubit. 
Columns are then included sequentially in increasing order, with the matrix rank checked after each addition; columns that do not increase the rank are discarded and added to the matrix $\M{H}_t$. 
This procedure is repeated until $\M{H}_s$ attains full rank, resulting in a square, invertible matrix, while the remaining matrix $\M{H}_t$ has dimensions $(n-r) \times r$.

Higher-order versions (\ac{OSD}-1, \ac{OSD}-2) refine the estimation by including qubits that were previously excluded from the matrix construction. This is done by setting $\hat{\mathbf{e}}_t$ to be different from $\mathbf{0}$ and computing the total error as~\cite{Rof20:BP}
\begin{align}
\label{OSD_higher}
    \hat{\mathbf{e}}_{[s,t]} = \left(\hat{\mathbf{e}}_{s} + \M{H}_{s}^{-1}\M{H}_{t}\hat{\mathbf{e}}_{t},\; \hat{\mathbf{e}}_{t}\right).
\end{align}
For \ac{OSD}-1, \eqref{OSD_higher} is evaluated for all weight-one configurations of $\hat{\mathbf{e}}_{t}$.  
For \ac{OSD}-2, \eqref{OSD_higher} is evaluated for all weight-two configurations restricted to the first $\lambda$ bits of $\hat{\mathbf{e}}_{t}$, where $\lambda$ is a user-defined parameter.
Among all possible solutions, the one with the lowest Hamming weight is selected.

\subsection{Belief Propagation based on Guided Decimation }

As a variant of the \ac{BP} algorithm, \ac{BPGD} was proposed in~\cite{Yao24:GD}.
This decoder executes successive \ac{BP} instances, where the subsequent ones adopt as input \acp{LLR} the output \acp{LLR} of the previous ones, except for the $j$-th \ac{LLR} corresponding to the largest absolute value which is set accordingly to
\begin{align}
p_j^{\text{in}} =
\begin{cases}
\infty, & \text{if } p_j^{\text{out}} > 0, \\
-\infty, & \text{if } p_j^{\text{out}} \le 0.
\end{cases}
\end{align}
Each \ac{BP} instance run $T$ \ac{BP} iterations.
The algorithm terminates when a valid solution is found or when all qubits are decimated.

\subsection{Relay Belief Propagation}

The \ac{RelayBP} algorithm has been recently proposed, incorporating memory across successive \ac{BP} instances~\cite{Mul25:relay}. 
If the first \ac{BP} instance fails to converge, the subsequent instance utilizes the output \ac{LLR} probabilities from the previous run when computing the input channel probabilities as
\begin{align}
    p_\ell^{\text{in}} = (1 - \gamma)\,p_0^{\text{in}} + \gamma\,p_{(\ell-1)}^{\text{out}}
\end{align}
where $\ell$ denote the iteration step and $\gamma$ is drawn from a uniform distribution over the interval 
$\big[\gamma_c - \frac{\gamma_w}{2}, \, \gamma_c + \frac{\gamma_w}{2}\big]$,
with $\gamma_c$ and $\gamma_w$ as input parameters. 
These parameters can be adjusted according to the specific quantum code in order to optimize the decoding performance. In general, the procedure is repeated until $S$ candidate solutions are obtained or until the maximum number of \ac{BP} instances (or leg length) is reached. Among the obtained candidates, the solution with the minimal weight is then selected for error correction.
As suggested in \cite{Mul25:relay}, the first leg employs $T_1$ iterations, while all the subsequent legs run $T_2 \le T_1$ iterations. 

\section{Restart Belief}
\label{sec:Belief}

In this section, we describe our proposal for a general \ac{CSS} \ac{QLDPC} decoder.  
In the following, we detail the decoding procedure for $\M{Z}$-type Pauli errors.
The same strategy applies to $\M{X}$-type errors, with $\M{Z}$ and $\M{X}$ interchanged.
For the sake of clarity, the proposal is summarized in Algorithm~\ref{algo:belief}.

\input{Figures/Algo/RestartBP}

\begin{figure*}[t]
    \centering
    \includegraphics[width=\textwidth]{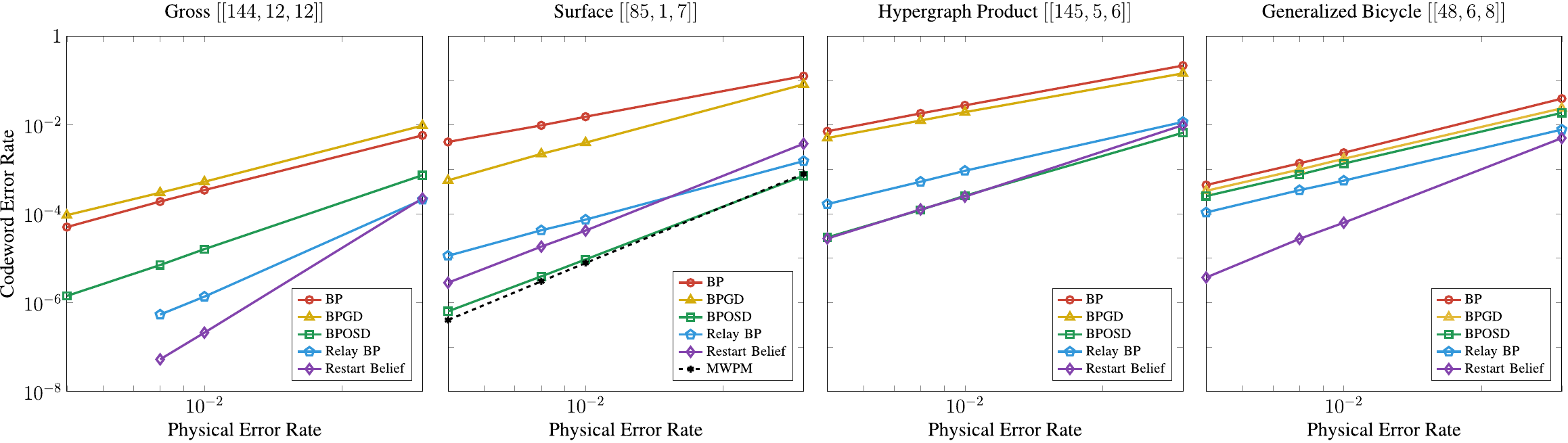}
    \caption{Codeword error rate vs. physical error rate varying codes and decoders.
    \ac{QLDPC} code classes from left to right: bivariate bicycle, topological, hypergraph product, and generalized bicycle. }
    \label{Fig:Res}
\end{figure*}

\subsection{Decoder Description}

As discussed in Section~\ref{sec:decodersLiterature}, the degeneracy of quantum codes often prevents the convergence of the \ac{BP} decoder. 
However, in most cases, we observe that the data qubits associated with the lowest \acp{LLR} correspond to the erroneous qubits.
For this reason, we start with an initial \ac{BP} instance, denoted in the following as the root \ac{BP}, executing $T_\mathrm{root}$ \ac{BP} iterations.
In Algorithm~\ref{algo:belief}, we employ the function $\texttt{beliefProp}(\V{p}^{\text{in}}, T)$, implementing a \ac{BP} instance with input \acp{LLR} $\V{p}^{\text{in}}$, $T$ \ac{BP} iterations, and returning the output \acp{LLR} $\V{p}^{\text{out}}$ along with the estimated error vector $\V{\hat{e}}$. 
Note that, if the \ac{BP} decoder does not converge, the estimated error vector is set to all zeros.
At this point, we check for early termination.
If $\V{\hat{e}}$ is consistent with the original syndrome, we found a valid solution.
However, this could not be the \ac{MW} solution.
For this reason, we add a distance-specific and a defect-specific early termination rule.
With the distance-specific rule we check if $\texttt{w}(\V{\hat{e}}) \leq t$ using the weight function $\texttt{w}$.
If this is the case we are sure to have found the \ac{MW} solution.
Conversely, the defect-specific condition triggers early termination if decoder detects that the error weight exceeds $t$. 
Specifically, we define as $\xi$ the maximum column weight.
In other words, $\xi$ represents the maximum number of $\M{X}$ parity checks that can anticommute per $\M{Z}$ error.
Hence, if $\texttt{w}(\V{s}) / \xi > t$, we can conclude with certainty that the original error has weight greater than $t$.
In this case, even if the weight of the obtained solution exceeds $t$, we accept it as the final solution.

In case $\V{\hat{e}}$ is not consistent with the original syndrome or none of the termination criteria are satisfied, the algorithm proceeds by sorting the output \acp{LLR} in ascending order.
The set $\mathcal{I}$ contains the indices of the reordered \acp{LLR} using the function $\texttt{sort}$.
Then, we start $\eta \le n$ parallelizable searches.
At the start of the $i$-th search (or branch) we apply a $\M{Z}$ error on the qubit with the $i$-th lowest output \ac{LLR}.
The temporary error guess of the $i$-th search is stored in $\V{\bar{e}}$.
We accordingly update the syndrome as $\V{\hat{s}}=~\V{\hat{s}}~\oplus~\left( \M{H}_\mathrm{x} \V{\bar{e}} \right)$, where $\V{\hat{s}}$ is a copy of the original syndrome $\V{s}$.
In addition, using the function \texttt{fixLLR}, we set the input \acp{LLR} to $\infty$ in the location specified by $\V{\bar{e}}$.
Consequently, another \ac{BP} instance is executed with $T_\mathrm{branch}$ \ac{BP} iterations.
If this instance of \ac{BP} does not converge, a $\M{Z}$ error is imposed on the qubit with the lowest output \ac{LLR}.
The syndrome and the input \acp{LLR} are then updated accordingly, as described above, and a new instance of \ac{BP} is executed. 
This procedure is repeated iteratively until either the \ac{BP} converges or $t = \lfloor (d - 1)/2 \rfloor$ $\M{Z}$ errors have been inserted into $\V{\bar{e}}$.
Next, we check if the final estimated error $\V{\hat{e}} = \V{\hat{e}} \oplus \V{\bar{e}}$ is consistent with the original syndrome.
In case $\V{\hat{e}}$ is consistent, we also check the same conditions used for the root solution, storing the current solution if it has the lowest weight so far and none of the termination criteria are met.
In case $\V{\hat{e}}$ is not consistent or none of the termination criteria are satisfied, we proceed with the exploration of the next branch, until the maximum number of searches $\eta$ have been explored.
Finally, if none of the early termination conditions are ever met, the decoder outputs the stored solution.

Note that, the parameter $\eta$ can be viewed as a trade-off lever between performance and latency.
In the following, we fine-tune this parameter to ensure that our decoder can systematically correct all error patterns of weight up to $t$.
This systematic nature in our decoding strategy, which provides performance guarantees, is one of its key strengths compared to more stochastic approaches, such as \ac{RelayBP}.

\subsection{Complexity Analysis} \label{sec:complexity}

The \ac{RB} decoder performs a single root \ac{BP} instance and at most $\eta$ branch searches, where each branch consists of at most $t-1$ \ac{BP} instances.
The root instance runs $T_\mathrm{root}$ iterations, while each branch instance runs $T_\mathrm{branch}$ iterations.
Considering a single branch, we have a computational cost of $\mathcal{O}((t-1) T_\mathrm{branch} n) \approx \mathcal{O}(t T_\mathrm{branch} n)$, due to the fact each \ac{BP} iteration has a complexity of $\mathcal{O}(n)$.
Hence, considering also the root \ac{BP} instance, the overall worst-case complexity of the \ac{RB} decoder is $\mathcal{O}((\eta t T_\mathrm{branch}  + T_\mathrm{root}) n)$. 
This scaling is similar to that of conventional \ac{BP}-based iterative decoders, while remaining considerably lower than that of \ac{OSD}-0, which exhibits a cubic complexity of $\mathcal{O}(n^{3})$ \cite{PanKalOSD:21}. 
It is worth noting that the \ac{RB} decoder can be parallelized in hardware, since each of the $\eta$ outer rounds can be executed independently, reducing the effective worst-case complexity to $\mathcal{O}((t T_\mathrm{branch}  + T_\mathrm{root}) n)$.

\section{Numerical Results}\label{sec:NumRes}

\input{Figures/Table_iterations}\label{Tab:iter}

In this section, we compare the decoders discussed in Section~\ref{sec:decodersLiterature} with the proposed decoder described in Section~\ref{sec:Belief} through extensive Monte Carlo simulations.
The comparison is carried out considering four different \ac{QLDPC} codes to show also the decoder flexibility.
The set of codes used in this comparison comprises a bivariate bicycle code, a topological code, an hypergraph product code, and a generalized bicycle code, thereby enabling the exploration of a broad spectrum of \ac{QLDPC} structures.
In particular, as topological code we adopt the $[[85,1,7]]$ surface code~\cite{BraKit:98}.
As hypergraph product code we adopt the $[[145, 5, 6]]$ code constructed from an augmented classical repetition code~\cite{Rof20:BP}.
As a bivariate bicycle code we adopt the $[[144, 12, 12]]$ gross code~\cite{Bra24:Gross}, recently used also in~\cite{Yod25:gross}.
As a generalized bicycle code we adopt the $[[48, 6, 8]]$ code from~\cite{KovPry:13}. 
For each physical error rate, the simulations are executed until $100$ decoding failures are observed, ensuring statistically meaningful estimates of the codeword error rate.
The source code implementation is available in our repository QBelief~\cite{QBelief25}.

For the Relay decoder, we use a leg length of $301$, $T_1 = 80$, $T_2 = 60$, $S = 5$, and $\alpha = 1$, as provided in~\cite{Mul25:relay}.
Also, we tested some values among the ones provided in~\cite{Mul25:relay} for $\gamma_c$ and $\gamma_w$, and select the pair $[-0.24,\,0.66]$ as the best-performing one.
For all the remaining decoders, we use $\alpha = 1 - 2^{N_\mathrm{iter}}$, where $N_\mathrm{iter}$ is the current iteration~\cite{CheFos02:alpha, Rof20:BP}.
For the \ac{BP}+\ac{OSD}, we employ an \ac{OSD}-2 with $\lambda = 10$ as suggested in~\cite{Rof20:BP}.
For the \ac{BP}+\ac{OSD}, the \ac{BPGD} and the baseline \ac{BP}, we set $T = 50$.
Also, for the \ac{BPGD} decoder we set $p_j^{\text{in}} = 25$ during decimation, as suggested in \cite{Yao24:GD}.
For the \ac{RB} decoder we set $T_\mathrm{root} = 50$ and $T_\mathrm{branch} = 10$.
The parameter $\eta$ has been fine-tuned for each code variant to guarantee the error correction capability $t$ for the adopted $\alpha$, $T_\mathrm{root}$, and $T_\mathrm{branch}$.
In particular, $\eta$ values are listed in Table~\ref{tab:times} for each code.

In Fig.~\ref{Fig:Res}, we report the codeword error rate of several code-decoder pairs over a depolarizing channel versus the depolarizing channel physical error rate.
From the figure, it can be observed that the \ac{RB} decoder outperforms the decoders based on \ac{BP}, \ac{BPGD}, and \ac{RelayBP}. 
Regarding the $[[85,1,7]]$ surface code, our proposal exhibits a performance gap compared to the \ac{BP}+\ac{OSD}.
As shown in~\cite{ForValChi24:MacW}, this constant gap is due to the fact that the \ac{RB} decoder correct less error pattern of weight $t+1$.
It is worth noting that the slope of the two curve is the same of the \ac{MWPM} decoder, meaning that both decoders protect the codeword within the code distance.
Regarding the gross code, we highlight that the \ac{RB} decoder outperforms \ac{RelayBP} on the gross code, while the \ac{BP}+\ac{OSD} is not able to preserve the code distance.
Regarding the $[[145, 5, 6]]$ hypergraph product code, the \ac{BP}+\ac{OSD}, the \ac{RelayBP}, and the \ac{RB} perform practically the same, with only a slight gap for the \ac{RelayBP}.
Finally, regarding the $[[48, 6, 8]]$ generalized bicycle, we observe a significant performance advantage when adopting the \ac{RB} decoder.

Table~\ref{tab:times} reports the average \ac{BP} iterations used by the two decoders offering the best compromise between hardware complexity and performance, the \ac{RB} and the \ac{RelayBP}. 
The table is generated by varying $n_\mathrm{e}$, the number of introduced $\M{Z}$ errors, to illustrate how computational cost depends on the error weight handled by the decoder.
The $\eta$ parameters listed in Table~\ref{tab:times} have been selected as the minimum value that guarantees the error correction capability.
In fact, we have exhaustively verified through simulations that our decoder successfully corrects all error patterns up to weight $t$ for all the listed codes.  
On the other hand, the non-deterministic nature of the \ac{RelayBP} leads to variability in the decoding outcomes, causing certain error patterns within the code distance to be occasionally misdecoded.
For all codes except to the gross code, the \ac{RB} decoder consistently proves to be faster than the \ac{RelayBP} decoder.
In the case of the gross code,
the \ac{RB} decoder requires a higher number of \ac{BP} iterations for errors of weight $n_\mathrm{e} = t + 1$ and $n_\mathrm{e} = t + 2$.
This is because the defect-specific criterion cannot uniquely identify such error patterns. 

The selected set of codes provides a solid benchmark for evaluating decoder performance, owing to their diverse decoder performance trends, computational requirements, and \ac{QLDPC} structure.


\section{Conclusions}\label{sec:conclusions}

In this work, we introduce a novel decoder for \ac{QLDPC} codes. Across all tested quantum codes, our results show that the proposed \ac{RB} decoder achieves the lowest logical error rate while maintaining a computational complexity comparable to other \ac{BP}-based decoders reported in the literature. 
As additional features, we highlight that the \ac{RB} decoder is parallelizable, easy-to-tune, and deterministic.
The simulation analyses, complemented with exhaustive searches, show that our decoder guarantees the error correction capability for all the tested codes.


\bibliographystyle{IEEEtran}
\bibliography{Files/IEEEabrv,Files/StringDefinitions,Files/StringDefinitions2,Files/refs}

\end{document}

%% file: Files/Acronimi_SICMMA.tex
\begin{acronym}
\small
\acro{ASIC}{application-specific integrated circuit}
\acro{AWGN}{additive white Gaussian noise}
\acro{BB}{bivariate bicycle}
\acro{BC}{bubble clustering}
\acro{BCH}{Bose–Chaudhuri–Hocquenghem}
\acro{BD}{bounded distance}
\acro{BP}{belief propagation}
\acro{BPGD}{belief propagation based on guided decimation}
\acro{CDF}{cumulative distribution function}
\acro{CER}{codeword error rate}
\acro{CRC}{cyclic redundancy code}
\acro{CSS}{Calderbank, Shor, and Steane}
\acro{FPGA}{field-programmable gate array}
\acro{i.i.d.}{independent identically distributed}
\acro{LDPC}{low-density parity-check}
\acro{LEMON}{library for efficient modeling and optimization in networks}
\acro{LLR}{log-likelihood ratio}
\acro{LUT}{lookup table}
\acro{ML}{maximum likelihood}
\acro{MPS}{matrix product state}
\acro{MST}{minimum spanning tree}
\acro{MW}{minimum weight}
\acro{MWPM}{minimum weight perfect matching}
\acro{OSD}{ordered statistics decoding}
\acro{PDF}{probability density function}
\acro{PMF}{probability mass function}
\acro{PruST}{pruned spanning tree}
\acro{QEC}{quantum error correction}
\acro{QECC}{quantum error correcting code}
\acro{QLDPC}{quantum low-density parity-check}
\acro{RFire}{Rapid-Fire}
\acro{RelayBP}{relay belief propagation}
\acro{RB}{restart belief}
\acro{STM}{spanning tree matching}
\acro{UF}{union-find}
\acro{WE}{weight enumerator}
\acro{WEP}{weight enumerator polynomial}
\end{acronym}

%% file: Figures/Algo/RestartBP.tex
\begin{algorithm}[t]
\SetKwInOut{Input}{input}
\SetKwInOut{Output}{output}
\caption{Restart Belief} \label{algo:belief}
\Input{$t$, $\xi$, $\V{s}$,  $\M{H}_\mathrm{x}$; \\
} 
\Output{$\V{\hat{e}}$, estimated error; \\  } 
\BlankLine
$\V{p}^{\text{in}} \gets \log \left ( (1 - p) / p \right )$; \\
$[\V{\hat{e}},\V{p}^{\text{out}}] \gets \texttt{beliefProp}(\V{p}^{\text{in}}, T_\mathrm{root})$; \\
\If{$\M{H}_\mathrm{x} \V{\hat{e}} = \V{s}$}{ \If{$\texttt{w}(\V{\hat{e}}) \leq t$ or $\texttt{w}(\V{s}) /  \xi > t$}{
\Return $\V{\hat{e}}$; \\
}
}
$\mathcal{I} \gets \texttt{sort}
(\V{p}^{\text{out}})$; \\
$w_\mathrm{min} \gets \infty$; $\V{\hat{s}} \gets \V{s}$; \\
\ForAll{$i \in 1,\dots,\eta$}{
$\V{p}^{\text{in}} \gets \log (  (1 - p) / p  )$; \\
$\V{\bar{e}} \gets 0$; \\
$\V{\bar{e}}(\mathcal{I}[i]) \gets 1$; \\
\ForAll{$j \in 1, \dots, t - 1$}{
$\V{\hat{s}} \gets \V{\hat{s}} \oplus \left( \M{H}_\mathrm{x} \V{\bar{e}} \right)$; \\
$\V{p}^{\text{in}} \gets \texttt{fixLLR}(\V{p}^{\text{in}}, \V{\bar{e}})$; \\
$[\V{\hat{e}},\V{p}^{\text{out}}]\gets \texttt{beliefProp}(\V{p}^{\text{in}}, T_\mathrm{branch})$; \\
\InlineIfThen{$\M{H}_\mathrm{x} \V{\hat{e}} = \V{\hat{s}}$}{
\textbf{break}; 
}\\
$\V{\bar{e}}[\texttt{argmin}(\V{p}^{\text{out}})] \gets 1$; \\
}
$\V{\hat{e}} \gets \V{\hat{e}} \oplus \V{\bar{e}}$; \\
\If{$\M{H}_\mathrm{x} \V{\hat{e}} = \V{s}$}{ 
\InlineIfThen{$\texttt{w}(\V{\hat{e}}) \leq t$ or $\texttt{w}(\V{s}) /  \xi > t$}{
\Return $\V{\hat{e}}$; \\
}
\If{$\texttt{w}(\V{\hat{e}}) < w_\mathrm{min}$}{
$\V{e}_\mathrm{min} \gets \V{\hat{e}}$; \, $w_\mathrm{min} \gets \texttt{w}(\V{\hat{e}})$; \\
}
}
}
$\V{\hat{e}} \gets \V{e}_\mathrm{min}$; \\
\Return $\V{\hat{e}}$; \\
\end{algorithm}

%% file: Figures/Table_iterations.tex
\begin{table*}[t]
    \centering
    \setlength{\tabcolsep}{3pt}
    \caption{Average BP Iterations}
    \label{tab:times}
    \resizebox{\textwidth}{!}{
    \begin{tabular}{cC{2.1cm}C{2cm}C{1.5cm}C{1.5cm}C{1.5cm}C{1.5cm}C{1.5cm}C{1.5cm}C{1.5cm}C{1.5cm}C{1.5cm}C{1.5cm}C{1.5cm}C{1.5cm}}
        \toprule 
        \rowcolor[gray]{.95}
        Code & Decoder & Configuration & $n_\mathrm{e}=1$ & $n_\mathrm{e}=2$ & $n_\mathrm{e}=3$ & $n_\mathrm{e}=4$ & $n_\mathrm{e}=5$ & $n_\mathrm{e}=6$ & $n_\mathrm{e}=7$ & $n_\mathrm{e}=8$ & $n_\mathrm{e}=9$ & $n_\mathrm{e}=10$ & $n_\mathrm{e}=11$ & $n_\mathrm{e}=12$ \\
        \midrule
        & \textbf{Restart Belief} & $\eta = 48$ & 1.000 & 5.085 & 35.44 & 812.3 & 792.9 & 859.1 & 799.7 & 842.9 \\[-11pt]
        \multirow{-2}{*}[-10pt]{$[[48, 6, 8]]$} \\[-1pt]
        & \textbf{Relay BP} & & 91.66 & 191.8 & 478.3 & 3361 & 13831 & 14458 & 14633 & 14848 \\
        \midrule
        & \textbf{Restart Belief} & $\eta = 35$ & 1.000 & 1.102 & 1.577 & 2.545 & 3.249 & 370.7 & 275.3 & 64.57 & 66.22 & 39.22 & 77.50 & 138.6 \\[-11pt]
        \multirow{-2}{*}[-10pt]{$[[144,12,12]]$} \\[-1pt]
        & \textbf{Relay BP} &  & 56.28 & 70.49 & 91.09 & 111.4 & 133.5 & 178.3 & 242.1 & 291.1 & 372.9 & 531.9 & 972.6 & 2065 \\
        \midrule
        & \textbf{Restart Belief} & $\eta = 8$ & 2.000 & 4.628 & 11.03 & 41.36 & 50.65 & 72.99 & 91.75 & 109.2 & & & & \\[-11pt]
        \multirow{-2}{*}[-10pt]{$[[85,1,7]]$} \\[-1pt]
        & \textbf{Relay BP} &  & 31.42 & 67.52 & 145.3 & 222.1 & 341.1 & 445.1 & 585.0 & 738.1 & & & & \\
        \midrule
        & \textbf{Restart Belief} & $\eta = 6$ & 1.891 & 3.484 & 24.31 & 15.52 & 22.84 & 33.33 & & & & & & \\[-11pt]
        \multirow{-2}{*}[-10pt]{$[[145,5,6]]$} \\[-1pt]
        & \textbf{Relay BP} &  & 45.47 & 69.78 & 108.8 & 158.8 & 203.4 & 243.1 & & & & & & \\
        \bottomrule
    \end{tabular}}
\end{table*}